\documentclass[12pt]{JHEP3}

\preprint{  }

\usepackage{epsfig,multicol}
\usepackage{amsmath}
\usepackage{graphics}
\usepackage{graphicx}
\usepackage{dcolumn}
\usepackage{amssymb}
\usepackage{amsthm}
\usepackage{amsfonts}
\usepackage{subfigure}
\usepackage{setspace}

\title{Detect black holes using photons for coupling model of electromagnetic and gravitational fields}

\author{ Jiliang  Jing\footnote{Electronic address:
jljing@hunnu.edu.cn}, Songbai Chen, Qiyuan Pan,  Jieci Wang \\ 
Department of Physics, Key Laboratory of Low Dimensional Quantum Structures and Quantum Control of Ministry of Education, and Synergetic Innovation Center for Quantum Effects and Applications, Hunan Normal University, Changsha, Hunan 410081, P. R. China. 
}

\abstract{
For a model of the electromagnetic field coupled to Weyl tensor, Maxwell equations are modified and  photons at low frequencies no longer propagate along light cone. If we detect a black hole using these photons,  some difficulties appear because we can not determine the position of event horizon which is defined by null surface.  To overcome these difficulties, the simplest way may be an effective description by introducing an effective spacetime in which the photons propagate along the light cone. Then, we find,  comparing the results with those of the original spacetime, that the event horizon and temperature do not change, but the area of the event horizon and Bekenstein-Hawking entropy become different. We show that the total entropy for this system, which is still the same as that of original spacetime, consists of two parts, one is the Bekenstein-Hawking entropy and the other is the entropy arising from the coupling of electromagnetic field and Weyl tensor. We also present the effective descriptions for the Smarr relation and first law of thermodynamics.
}


\begin{document}

\section{Introduction}

The studies of coupling models of the electromagnetic and
gravitational fields have attracted much attention in many fields
recently \cite{Drummond, Balakin, Balakin1,Faraoni, Hehl,
Balakin05,Bamba,Balakin09} due to the facts: (a) the interaction
between the electromagnetic and gravitational fields could be
appeared naturally in quantum electrodynamics in curved spacetime
\cite{Drummond},  and (b) the Maxwell and Einstein equations
modified by the coupling terms in the Lagrangian will lead to
emergence of new properties of the gravitational and electromagnetic
fields. Inspired by a well motivated non-minimal coupling between
gravity and electromagnetism, Balakin and Zimdahl \cite {Balakin05}
demonstrated that the richer structure of the corresponding theory
gives rise to novel features of the cosmological dynamics. Bamba and
Nojiri \cite{Bamba} reviewed cosmology in non-minimal Maxwell theory
in which the electromagnetic field couples to a function of the
scalar curvature of the spacetime and explored the forms of the
non-minimal gravitational coupling which generate the finite-time
future singularities and the general conditions for this coupling in
order that the finite-time future singularities cannot appear.
Balakin and Muharlyamov \cite{Balakin09} showed that the resonance
interactions between particles and transversal waves in plasma can
take place due to the curvature coupling effect  in a coupling model
of gravity with electromagnetic fields. Drummond and Hathrell
\cite{Drummond} noted that the characteristics of propagation of the
light in the coupling models are altered by the tidal gravitational
forces on the photons, so that in some cases photons travel at
speeds greater than unity. Daniels, Cai, Shore, and Hollowood
\emph{et al.} \cite{Dan,Cai,Shore,Shore1,Hollowood,Hollowood2,Cho,Dalvit} proved that
the interactions between the electromagnetic field and the spacetime
curvature lead to a dependence of the photon velocity on the motion
and polarization directions. In these coupling models the
electromagnetic field is described by the modified theory rather
than the usual free Maxwell theory. Therefore, the characteristics
of propagation of the light are altered and the laws of geometric
optics, in general, are invalid in the original spacetime.  We
\cite{Jing Ann} showed that, by introducing an effective spacetime,
the wave vector can be cast into null and then obeys the null
geodesic equation, the polarization vector is perpendicular to the
rays, and the number of photons is conserved. We also found that the
focusing theorem of light rays in the effective spacetime can also
be written as the usual form.

Noting that photons at low frequencies no longer propagate along the
light cone in the coupling models of the electromagnetic and
gravitational fields, if we detect a black hole by using these
photons, we can not determine the event horizon of a black hole by
means of the null surface, and we can not calculate quantum entropy
of black hole using statistical mechanics because we do not have any
knowledge of the statistical mechanics if the speed of the light is
greater than unity. We also do not know whether the thermodynamical
laws still valid in  these models. It should be noted that all the
difficulties arise from the fact that photons no longer propagate
along the light cone in these coupling models. Therefore, to
overcome these difficulties, the simplest way may be to introduce an
effective spacetime in which the photons propagate along the light
cone.  In this manuscript we systematically present an effective
description for the thermodynamics in a coupling model of
electromagnetic and gravitational fields, and some interesting
results are obtained.

The plan of the paper is as follows. In the next section we
introduce the difficulties arise in coupling model of
electromagnetic and gravitational fields. In section III we will
study the temperature and entropy of black hole described by the
effective spacetime. In section IV we will obtain the effective
Smarr formula and first law of thermodynamics. We present our
conclusions and give some discussions in the last section.

\section{Difficulties for physics of black hole in coupling model of electromagnetic and gravitational fields}

The electromagnetic theory with the Weyl correction has been studied extensively \cite{Wu2011,Ma2011, Momeni,Roychowdhury,zhao2013,Weyl1}. The action for a model of the electromagnetic field coupled to the Weyl tensor can be taken as  \cite{Weyl1}
\begin{eqnarray}
S=\int d^4 x \sqrt{-g}\bigg[\frac{R}{16\pi
G}-\frac{1}{4}\bigg(F_{\mu\nu}F^{\mu\nu}-4\alpha
C^{\mu\nu\rho\sigma}F_{\mu\nu}F_{\rho\sigma}\bigg)\bigg], \label{acts}
\end{eqnarray}
where $F_{\mu\nu}=A_{\nu;\mu}-A_{\mu;\nu}$ is the usual
electromagnetic tensor with a vector potential $A_\mu$, $\alpha$ is
the coupling constant, and $C_{\mu\nu\rho\sigma}$ is the Weyl tensor of the spacetime. Varying the action
(\ref{acts}) with respect to vector potential $A_{\mu}$, we get
\begin{eqnarray}
\nabla_\mu(F^{\mu\nu}-4\alpha C^{\mu\nu\rho\sigma}F_{\rho
\sigma})=0. \label{main 1}
\end{eqnarray}
Under the geometric optics assumption in which the wavelength $\lambda$ of photon is much smaller than the minimum $L$ of  the radius of curvature of a wavefront $\mathfrak{L}$ and the radius of curvature of the spacetime $\mathfrak{R}$, the electromagnetic field strength can be written as
\begin{eqnarray}
F_{\mu\nu}=f_{\mu\nu}e^{i\Theta(t,r,\theta,\phi) },\label{ef1}
\end{eqnarray}
where $f_{\mu\nu}$ is a slowly varying amplitude, and
$\Theta(t,r,\theta,\phi)$ is a rapidly varying phase and the wave
vector can be defined as
$k_{\mu}=\partial_{\mu}\Theta(t,r,\theta,\phi)$. According to the
Bianchi identity, we note that
$f_{\mu\nu}=k_{\mu}a_{\nu}-k_{\nu}a_{\mu}$ in which $a_{\mu}$ is the
polarization vector satisfying the condition that
$k_{\mu}a^{\mu}=0$. Substituting Eq. (\ref{ef1}) into Eq. (\ref{main
1}), we get the equation of motion of photon coupling to Weyl tensor
\begin{eqnarray}
k_{\mu}k^{\mu}a^{\nu}+8\alpha
C^{\mu\nu\rho\sigma}k_{\sigma}k_{\mu}a_{\rho}=0.\label{main 3}
\end{eqnarray}
Eq. (\ref{main 3}) shows that the wave vector for this model is not
null vector. That is to say, the interactions between the
electromagnetic field and the spacetime curvature lead to a
dependence of the photon velocity on the motion and polarization
directions, so that in some cases photons travel at speeds greater
or less than unity.

If we use this signal to detect the black hole, the fact that
photons no longer propagate along the light cone will make the
physics of black hole more complicated for the following reasons: 1)
We can not get the position of the event horizon of a black hole by
means of the null surface  because the photons no longer propagate
along the light cone. 2) We do not have any knowledge of the
statistical mechanics if the speed of the light is greater than
unity but the usual method for calculating quantum entropy of black
hole is  based on the statistical mechanics.  3) We do not know
whether the thermodynamical laws are valid for the case that the
speed of the light is greater than unity.

It should be pointed out, in this coupling model, that all the
difficulties arise from the fact that photons no longer propagate
along the light cone. Therefore, to overcome these difficulties, the
simplest way may be to introduce an effective spacetime in which
photons propagate along the light cone. Therefore, we will find the
effective spacetime in next section.

\section{Effective spacetime for the coupling model}

In a general four-dimensional static and spherically symmetric
spacetime\begin{eqnarray} ds^2&=&g_{00} dt^2+g_{11} dr^2+g_{22}(
d\theta^2+sin^2\theta d\phi^2),\label{m1}
\end{eqnarray}
where $g_{00}$, $g_{11}$ and $g_{22}$ are functions of the polar
coordinate $r$ only,  by means of  $
g_{\mu\nu}=\eta_{ab}e^a_{\mu}e^b_{\nu}
$
where $\eta_{ab}$ is the Minkowski metric with the signature $(-,+,+,+)$,   for metric (\ref{m1})  we get the vierbeins
\begin{eqnarray}
e^a_{\mu}=diag(\sqrt{-g_{00} },\;\sqrt{g_{11} },\;\sqrt{g_{22} } ,\;\sqrt{g_{33} }).
\end{eqnarray}
Taking advantage of the antisymmetric combination of vierbeins \cite{Drummond}
\begin{eqnarray}
U^{ab}_{\mu\nu}=e^a_{\mu}e^b_{\nu}-e^a_{\nu}e^b_{\mu},
\end{eqnarray}
the Weyl tensor can be expressed as
\begin{eqnarray}
C_{\mu\nu\rho\sigma}=\mathcal{A}\bigg(2U^{01}_{\mu\nu}U^{01}_{\rho\sigma}-
U^{02}_{\mu\nu}U^{02}_{\rho\sigma}-U^{03}_{\mu\nu}U^{03}_{\rho\sigma}
+U^{12}_{\mu\nu}U^{12}_{\rho\sigma}+U^{13}_{\mu\nu}U^{13}_{\rho\sigma}-
2U^{23}_{\mu\nu}U^{23}_{\rho\sigma}\bigg),
\end{eqnarray}
with
\begin{eqnarray}
\mathcal{A}&=&-\frac{1}{12 (g_{00} g_{11} )^2 g_{22}
}\bigg\{\left[g_{00} g_{11} g_{00} ''-\frac{1}{2}(g_{00} g_{11}
)'g_{00} '\right]g_{22}-g_{00} ^2g_{11} g_{22} '' \nonumber \\
&&+\frac{1}{2}(g_{00} ^2g_{11} '-g_{00} g_{11} g_{00}
')g_{22}'-2(g_{00} g_{11} )^2\bigg\},
\end{eqnarray}
where the prime represents the derivative with respect to $r$. To simplify the equation of motion for the coupled photon propagation, we introduce three linear
combinations of momentum components \cite{Drummond}
\begin{eqnarray}
l_{\nu}=k^{\mu}U^{01}_{\mu\nu},~~~~~~~~
n_{\nu}=k^{\mu}U^{02}_{\mu\nu},~~~~~~~~
m_{\nu}=k^{\mu}U^{23}_{\mu\nu},
\end{eqnarray}
together with the dependent combinations
\begin{eqnarray}
&&p_{\nu}=k^{\mu}U^{12}_{\mu\nu}=\frac{1}{e^0_0 k^0}\bigg(e^1_1 k^1 n_{\nu}-e^2_2\tilde{k}^2 l_{\nu}\bigg),\nonumber\\
&&r_{\nu}=k^{\mu}U^{03}_{\mu\nu}=\frac{1}{e^2_2 \tilde{k}^2}\bigg(e^0_0 k^0 m_{\nu}+e^3_3k^3 l_{\nu}\bigg),\nonumber\\
&&q_{\nu}=k^{\mu}U^{13}_{\mu\nu}=\frac{e^1_1 k^1}{e^2_2 \tilde{k}^2}m_{\nu}+
\frac{e^1_1 e^3_3 k^1k^3}{e^0_0 e^2_2 k^0 \tilde{k}^2}n_{\nu}-\frac{e^3_3 k^3}{e^0_0 k^0}l_{\nu},\label{vect3}
\end{eqnarray}
where the vectors $l_{\nu}$, $n_{\nu}$, $m_{\nu}$ are independent and
orthogonal to the wave vector $k_{\nu}$. Contracting Eq. (\ref{main
3}) with $l_{\nu}$, $n_{\nu}$, $m_{\nu}$ respectively, using the
relation (\ref{vect3}) and introducing three independent
polarisation components $(a\cdot l)$, $(a\cdot n)$, and $(a\cdot
m)$, we find that the equation of motion of the photon coupling with
the Weyl tensor can be written as
\begin{eqnarray}
\left(\begin{array}{ccc}
K_{11}&0&0\\
K_{21}&K_{22}&
K_{23}\\
0&0&K_{33}
\end{array}\right)
\left(\begin{array}{c}
a \cdot l\\
a \cdot n
\\
a \cdot m
\end{array}\right)=0,\label{Kk}
\end{eqnarray}
with
\begin{eqnarray}
K_{11}&=&(1+16\alpha \mathcal{A})(g^{00}k_0k_0+g^{11}k_1k_1)+(1-8\alpha \mathcal{A})(g^{22}k_2k_2+g^{33}k_3k_3),\nonumber\\
K_{22}&=&(1-8\alpha \mathcal{A})(g^{00}k_0k_0+g^{11}k_1k_1+g^{22}k_2k_2+g^{33}k_3k_3),
\nonumber\\
K_{21}&=&24\alpha \mathcal{A} \sqrt{g^{11}g^{22}}k_1k_2,\;\;\;\;\;\;\;\;\;\;
K_{23}=8\alpha \mathcal{A}\sqrt{-g^{00}g^{33}}k_0k_3,\nonumber\\
K_{33}&=&(1-8\alpha \mathcal{A})(g^{00}k_0k_0+g^{11}k_1k_1)+(1+16\alpha \mathcal{A})(g^{22}k_2k_2+g^{33}k_3k_3). \nonumber \\
\end{eqnarray}
The condition of Eq. (\ref{Kk}) with the non-zero solution is
$K_{11}K_{22}K_{33}=0$. The first root $K_{11}=0$ leads to the
modified light cone
\begin{eqnarray}
(1+16\alpha \mathcal{A})(g^{00}k_0k_0+g^{11}k_1k_1)+(1-8\alpha \mathcal{A})(g^{22}k_2k_2+g^{33}k_3k_3)=0, \label{Kk31}
\end{eqnarray}
which corresponds to the case where the polarisation vector $a_{\mu}$ is proportional to $l_{\mu}$, i.e.,  the photon with the polarization along $l_{\mu}$. The second root $K_{22}=0$
corresponds to an unphysical polarisation and should be neglected.
The third root is $K_{33}=0$, i.e.,
\begin{eqnarray}
(1-8\alpha \mathcal{A})(g^{00}k_0k_0+g^{11}k_1k_1)+(1+16\alpha \mathcal{A})(g^{22}k_2k_2+g^{33}k_3k_3)=0,\label{Kk32}
\end{eqnarray}
which means that the vector $a_{\mu}=\lambda m_{\mu}$, i.e.,  the photon with the polarization along $m_{\mu}$.

Eqs. (\ref{Kk31}) and (\ref{Kk32}) show that the light cone
condition depends on not only the coupling between the photon and
the Weyl tensor, but also on the polarizations. We know that the
light cone condition is not modified for the radially directed
photons (i.e., $k_2 = k_3 =0$) but is modified for the orbital
photons (i.e., $k_1 =k_2 =0$), and the velocities of the photons for
the two polarizations are different,  i.e., the phenomenon of
gravitational birefringence. It is interesting to note that if
photons travel at speeds greater than unity for the case that the
photon travels with the polarization along $l_{\mu}$, then photons
will travel at speeds less than unity for the case that the photon
travels with the polarization along $m_{\mu}$, and vice versa.

Moreover, the light cone conditions (\ref{Kk31}) and (\ref{Kk32})
imply that the coupled photons propagation dose not along the  null
geodesic in the original metric (\ref{m1}). However, it is
interesting to note that we can cast the light cone conditions
(\ref{Kk31})  and (\ref{Kk32}) into null form
\begin{eqnarray}
\tilde{k}^\mu \tilde{k}_\mu=0,
\end{eqnarray}
 by  defining
$
\tilde{k}^\mu={\cal G}^{\mu\nu}k_\nu,~
\tilde{k}_\mu =k_\mu $  with effective contravariant metric ${\cal G}^{\mu\nu}$
\begin{eqnarray}
&&{\cal G}^{00}=g^{00},  \nonumber\\ &&  {\cal G}^{11}=g^{11}, \nonumber\\
&&{\cal G}^{22}= \left(\frac{1-8\alpha \mathcal{A}}{1+16\alpha \mathcal{A}}\right)^n g^{22}, \nonumber\\
&&{\cal G}^{33}= \left(\frac{1-8\alpha \mathcal{A}}{1+16\alpha \mathcal{A}}\right)^n g^{33},  \label{eff m3}
\end{eqnarray}
where $n=1$ for the case (\ref{Kk31}), and $n=-1$ for the case
(\ref{Kk32}). It should be noted that, in physics,  the metric
(\ref{eff m3})  with $n=1$ corresponds to the case where the photon
travels with the polarization along direction $l_{\mu}$, and the
metric with $n=-1$  to the case that the photon travels with the
polarization along direction $ m_{\mu}$.

\section{Effective temperature and entropy  of black hole }

We now study the event horizon, the area of event horizon, the temperature and entropy of black hole in the effective spacetime.

\subsection{Effective event horizon  of black hole}

Since photons propagate along the light cone in the effective spacetime, we can get the position of the event horizon of the black
hole by means of the null surface. For the static spherically symmetric spacetime described by effective metric  (\ref{eff m3}),
by using usual way that the event horizon should be a null surface $f(r)$ with normal vector $n_\mu=\frac{\partial
f(r)}{\partial x^\mu}$,  we know that the event horizon for the black hole can be obtained from the equation
\begin{eqnarray}{ \cal G}^{11}=g^{11}=0,
\end{eqnarray}
which shows that the event horizon  in the effective spacetime is the same as that of the original spacetime. Although the event
horizon does not change, the area of the event horizon is given by
\begin{eqnarray}\label{aera}
{\tilde{A}}_H=\int \sqrt{ {\cal G}_{22} {\cal G}_{33}}\Big{|}_{r_H}d\theta d\varphi=4 \pi \left[\left(\frac{1+16\alpha \mathcal{A}}{1-8\alpha \mathcal{A}}\right)^n g_{22}\right]_{r_H} ,
\end{eqnarray}
which is different from that of the original spacetime.  It is interesting to note that if the area of the event horizon for case
$n=1$ is greater than that of the original spacetime,  then the area  for case $n=-1$ will be less than that of the original spacetime.

\subsection{Effective temperature  of black hole}

A semi-classical method of modeling Hawking radiation as a tunnelling effect has been developed and has attracted a lot of interest \cite{man,wil,ag,sh3,par,arz,ding,ding2}. Tunnelling provides not only a useful verification of thermodynamic properties of black holes but also an alternate conceptual means for understanding the underlying physical process of black hole radiation.

Noting that the metric (\ref{eff m3})  take the same form as Eq.
(2.1) in Ref. \cite{ding3} in which the Hawking radiation was
studied by Hamilton-Jacobi ansatz method,  we know that  the Hawking
temperature  of the effective spacetime is described by
\begin{eqnarray} \label{TT1}
\tilde{T}= \frac{1}{4\pi } \left(\frac{1}{\sqrt{ {\cal G}_{00} {\cal G}_{11}}} \frac{d {\cal G}_{00}}{dr}\right)_{r_H}=\frac{1}{4 \pi }
\left(\frac{1}{\sqrt{g_{00}g_{11}}}\frac{d g_{00}}{dr}\right)_{r_H}=T,
\end{eqnarray}
which shows that the Hawking temperature $\tilde{T}$ of the effective spacetime is the same with the temperature $T$ of the
original spacetime.

\subsection{Effective entropy of black hole }

The statistical-mechanical entropy of black hole can be derived from
the canonical formulation \cite{Frolov98}. The corresponding free
energy, $F$, can be defined in term of the one-particle spectrum.
One of the ways to calculate $F$ is ``brick wall" model (BWM)
proposed by 't Hooft \cite{Hooft85} which has been used extensively
\cite{jing,agh,rbm,jlj,Jil}. In this model, in order to eliminate
divergence which appears due to the infinite growth of the density
of states close to the horizon, 't Hooft introduces a ``brick wall"
cutoff. We now use this approach to find the entropy of the
effective spacetime.

For the effective black hole, taking the  phase as
\begin{equation}
\Theta(t,r,\theta,\phi) =exp[-iEt+im\varphi+iW(r,\theta)], \label{phi}
\end{equation}
and using the equation of null wave vector $\tilde{k}_\mu
\tilde{k}^\mu=0$, we get
\begin{equation}
\tilde{k}_r^2=-\frac{{\cal G}_{rr}}{{\cal G}_{tt}} \left[E^2+{\cal G}_{tt}\left(\frac{m^2}{{\cal G}_{\varphi\varphi}}+\frac{\tilde{k}_\theta^2}
{{\cal G}_{\theta\theta}}\right)\right].\label{pr}
\end{equation} Therefore, the  number of modes with $E$, $m$ and
$\tilde{k}_{\theta}$ in phase space is shown by \cite{Padmanabhan86}
\begin{equation}
n(E, m, \tilde{k}_{\theta})=\frac{1}{\pi} \int d\theta
\int^{r}_{r_H+h} \tilde{k}_r dr=\frac{1}{\pi} \int d\theta
\int^{r}_{r_H+h}\sqrt{-\frac{{\cal G}_{rr}}{{\cal G}_{tt}} \left[E^2+{\cal G}_{tt}\left(\frac{m^2}{{\cal G}_{\varphi\varphi}}+\frac{\tilde{k}_\theta^2}
{{\cal G}_{\theta\theta}}\right)\right]}dr.\label{sqrt}
\end{equation}
The free  energy is given by
\begin{eqnarray}
\beta F&=& \int dm \int d\tilde{k}_{\theta}\int dn(E, m,
\tilde{k}_{\theta})ln \left[ 1-e^{-\beta E}\right] \nonumber \\
&=&-\beta \int dm \int d\tilde{k}_{\theta}\int \frac{n(E, m,
\tilde{k}_{\theta})} {e^{\beta E}-1} dE. \label{f1}
\end{eqnarray}
Substituting Eq.  (\ref{sqrt}) into Eq.  (\ref{f1}) and then taking
the integration over $E$ , we can work out the free energy
\begin{eqnarray}
F &=&\frac{1}{192\pi}\frac{\beta_H^3}{\beta ^4 \epsilon^2}\int d\theta  \left\{ \left(\frac{1+16\alpha \mathcal{A}}{1-8\alpha \mathcal{A}}\right)^n\sqrt{{g}_{\theta \theta}{g}_{\varphi
\varphi}}\right\}_{r_H}
+\frac{1}{1440}\frac{\beta_H^3}{\beta^4} \int
d\theta  \left\{ \left(\frac{1+16\alpha \mathcal{A}}{1-8\alpha \mathcal{A}}\right)^n\sqrt{{g}_{\theta \theta}{g}_{\varphi \varphi}} \right.
\nonumber \\
& &\times \left. \left[\frac{\partial
^2{g}^{rr}}{\partial r^2}+ \frac{3}{2}\frac{\partial
{g}^{rr}}{\partial r}\frac{\partial \ln |{g}_{tt}{g}_{rr}|}{\partial
r}+\frac{2\pi}{\beta_H \sqrt{-{g}_{tt}{g}_{rr}}}\frac{1-8\alpha \mathcal{A}}{1+16\alpha \mathcal{A}}\left(\frac{1}{{g}_{\theta \theta}}\frac{\partial }{\partial r}\frac{{g}_{\theta \theta}(1+16\alpha \mathcal{A})}{1-8\alpha \mathcal{A}}\right.\right.\right. \nonumber \\ &&
\left.\left.\left.
+\frac{1}{{g}_{\varphi \varphi}}\frac{\partial }{\partial r}\frac{{g}_{\varphi \varphi}(1+16\alpha \mathcal{A})}{1-8\alpha \mathcal{A}}\right)\right]\right\}_{r_H}\ln\frac{\Lambda}{\epsilon} ,  \label{f-0}
\end{eqnarray}
where $\delta^2=\frac{2\epsilon^2}{15}$ and
$\Lambda^2=\frac{L\epsilon ^2}{h}$   ( $\delta\approx
2\sqrt{h/\left(\frac{\partial {g}^{rr}}{\partial r}\right)_{r_H}}$
is the proper distance from the horizon to $\Sigma_h $, $\epsilon$
is the ultraviolet cutoff, and $\Lambda$ is the infrared cutoff).
Using the relation between  the entropy and the free energy, $
S=\beta^2\frac{\partial F}{\partial \beta}$, we find that the
leading term of the entropy is given by
\begin{eqnarray}
S&=&\frac{ { \tilde{A}}_{H}}{48\pi\epsilon^2}, \nonumber \label{smu}
\end{eqnarray}
where ${ \tilde{A}}_H$ is area of the event horizon defined by (\ref{aera}). Therefore, renormalizing the entropy by using the
standard approach shown in Refs. \cite{Birrell82,Solodukhin952,Jing2000}, we find that the Bekenstein-Hawking entropy of the effective black hole can be expressed as
\begin{eqnarray}
{\tilde{S}_H}=\frac{{\tilde{A}_H}}{4}= \pi \left[\left(\frac{1+16\alpha \mathcal{A}}{1-8\alpha \mathcal{A}}\right)^n g_{22}\right]_{r_H} ,
\end{eqnarray}
which  is different from that of the original  black hole. It is interesting to note that if the Bekenstein-Hawking entropy for case
$n=1$ is greater than that of the original  black hole,  the Bekenstein-Hawking entropy  for case $n=-1$ will be less than that
of the original  black hole.

\section{Effective Smarr formula and the first law of thermodynamics}

Since the Time-like Killing vector for the  metric  (\ref{eff m3}) is given by $\tilde{\xi}(t) ^\mu=(1,0,0,0)$,  we can take the mass as
\begin{eqnarray}\label{mass}
\tilde{M}&=&\frac{1}{4\pi}\int \tilde{R}^\mu_\nu \tilde{\xi}(t) ^\nu d\tilde{\Sigma}_\mu+ \frac{1}{4\pi}\int_{\partial S_B} \tilde{\xi}(t) ^{\mu;\nu}d\tilde{\Sigma}_{\mu\nu}, \nonumber \\
&=& \frac{1}{4\pi}\int \tilde{R}^\mu_\nu \tilde{\xi}(t) ^\nu d\tilde{\Sigma}_\mu+2 \tilde{T}\tilde{S},
\end{eqnarray}
where $\partial S_B$ is the boundary of the black hole. Because the value of the integral in Eq. (\ref{mass}) will be different for different black hole, we now show the Smarr formula and first law of thermodynamics for the effective Schwarzschild and Reissner-Nordstr\"{o}m  black holes, respectively.

\subsection{Effective Smarr formula and the first law of thermodynamics of  Schwarzschild black hole}

For the effective Schwarzschild black hole
\begin{eqnarray}
dS^2=-\left(1-\frac{r_+}{r}\right)dt^2+\left(1-\frac{r_+}{r}\right)^{-1}
dr^2+r^2\left(\frac{r^3-16 \alpha r_+}{r^3+8 \alpha
r_+}\right)^n(d\theta^2+sin^2\theta d\varphi^2),
 \end{eqnarray}
where $r_+=2 M$ is the radius of event horizon, since the integral
term in  Eq. (\ref{mass})  can be expressed as
\begin{eqnarray} \frac{1}{4\pi}\int \tilde{R}^\mu_\nu \tilde{\xi}(t) ^\nu d\tilde{\Sigma}_\mu=2 \tilde{T}\tilde{S}_C=2 T\tilde{S}_C,
\end{eqnarray}
we can obtain the Smarr relation and first law of the thermodynamics
\begin{eqnarray}\label{SCH 1}
M&=& 2 T (\tilde{S}_H+ \tilde{S}_C),  \\
d M&=&  T  (d \tilde{S}_H+ d \tilde{S}_C),
\end{eqnarray}
where  $\tilde{S}_H$ is the Bekenstein-Hawking  entropy and the
entropy $\tilde{S}_C$ can be considered as a contribution arising
from the coupling of electromagnetic field and the Weyl tensor,
which are given by
\begin{eqnarray}\label{SCH 2}
\tilde{S}_H&=&\left \{
\begin{array}{c}
 \pi r_+^2 \left(\frac{ r_+^2-16 \alpha }{r_+^2+8\alpha}\right), ~~~~~~(for ~ ~n=1),~  \\
 \pi r_+^2 \left(\frac{r_+^2+8 \alpha }{r_+^2-16\alpha}\right), ~~~~~(for ~ n=-1),\\
  \end{array}
 \right.   \\
 \tilde{S}_C&=& \left \{
\begin{array}{c}
~~ \frac{24 \pi \alpha r_+^2}{r_+^2+8\alpha}, ~~~~~~~~~~~~~~~(for ~ ~n=1),~  \\
  -\frac{24\pi \alpha r_+^2}{r_+^2-16\alpha}, ~~~~~~~~~~~~~~(for ~ n=-1). \\
\end{array}
 \right.
 \label{4dL}
\end{eqnarray}
The Smarr relation and first law of the thermodynamics show that the
entropy of the system  consists of two parts, $\tilde{S}_H$ and
$\tilde{S}_C$, and although  $ \tilde{S}_H$ and $ \tilde{S}_C$ are
different for the cases of $n=1$ and $n=-1$, the total entropy is
the same, i.e., $ \tilde{S}_H+ \tilde{S}_C=\pi r_+^2$ for the both
cases.

\subsection{Effective Smarr formula and the first law of thermodynamics of  Reissner-Nordstr\"{o}m black hole}

The effective Reissner-Nordstr\"{o}m black hole is described by
\begin{eqnarray}
dS^2&=&-\frac{(r-r_+)(r-r_-)}{r^2}dt^2+\frac{r^2}{(r-r_+)(r-r_-)}dr^2\nonumber
\\&+&r^2\left[\frac{r^4+32 \alpha r_+r_--16 \alpha r(r_++r_-)}{r^4-16
\alpha r_+r_-+8 \alpha r(r_++r_-)}\right]^n(d\theta^2+sin^2\theta
d\varphi^2),
 \end{eqnarray}
where $r_+=M+\sqrt{M^2-Q^2}$ and $r_-=M-\sqrt{M^2-Q^2}$. Noting that
\begin{eqnarray} \frac{1}{4\pi}\int \tilde{R}^\mu_\nu \tilde{\xi}(t) ^\nu d\tilde{\Sigma}_\mu=2 \tilde{T}\tilde{S}_C+\tilde{\Phi}_Q Q=2 T\tilde{S}_C+\tilde{\Phi}_Q Q,
\end{eqnarray}
from  Eq. (\ref{mass} ) we can obtain the following Smarr relation
and the first law of the thermodynamics
\begin{eqnarray}\label{RN 1}
M&=& 2 T (\tilde{S}_H+ \tilde{S}_C)+ \tilde{\Phi}_Q Q,  \\
d M&=&  T  (d \tilde{S}_H+ d \tilde{S}_C)+ \tilde{\Phi}_Q dQ,
\end{eqnarray}
where $Q$ is the electric charge and $\tilde{\Phi}_Q$ is the
corresponding potential of the charge, $\tilde{S}_H$ is the
Bekenstein  entropy  and the entropy $\tilde{S}_C$ can be considered
as a  contribution arising from the coupling of electromagnetic
field and the Weyl tensor, which are given by
\begin{eqnarray}\label{RN 2}
\tilde{\Phi}&=&\frac{Q}{r_+}\\
\tilde{S}_H&=&\left \{
\begin{array}{c}
 \pi r_+^2 \left[\frac{ r_+^3-16 \alpha (r_+-r_-)}{r_+^3+8\alpha (r_+-r_-)}\right], ~~~~~~(for ~ ~n=1),~  \\
 \pi r_+^2 \left[\frac{r_+^3+8 \alpha (r_+-r_-) }{r_+^3-16\alpha (r_+-r_-)}\right], ~~~~~(for ~ n=-1).\\
  \end{array}
 \right.   \\
 \tilde{S}_C&=& \left \{
\begin{array}{c}
~~ \frac{24\pi \alpha r_+^2(r_+-r_-)}{r_+^3+8\alpha(r_+-r_-)}, ~~~~~~~~~~~~~~(for ~ ~n=1),~ \\
- \frac{24\pi \alpha r_+^2(r_+-r_-)}{r_+^3-16\alpha(r_+-r_-)}, ~~~~~~~~~~~~~(for ~ n=-1).\\
\end{array}
 \right.
 \label{4dLL}
\end{eqnarray}
Although  $ \tilde{S}_H$  and $ \tilde{S}_C$ for the case of $n=1$
are different from those for the case of  $n=-1$, the total entropy
is also the same for the both cases, i.e., $ \tilde{S}_H+
\tilde{S}_C=\pi r_+^2$.

From above discussions we can obtain two interesting results: 1) The
entropy for this spacetime  consists of two parts, one is  the
Bekenstein-Hawking entropy, the other is the entropy arising from
the coupling of electromagnetic field and the Weyl tensor.  2)
Although the Bekenstein-Hawking  entropy $ \tilde{S}_H$  and the
entropy $ \tilde{S}_C$ arising from the coupling of electromagnetic
field and the Weyl tensor are different for the cases of $n=1$ and
$n=-1$, the total entropy is the same, i.e., $ \tilde{S}_H+
\tilde{S}_C=\pi r_H^2$ for the both cases.  That is to say, if the
entropy $ \tilde{S}_C$ arising from the coupling of electromagnetic
field and the Weyl tensor increases, the Bekenstein-Hawking  entropy
$ \tilde{S}_H$  will decrease, and vice versa.

\section{Conclusions and discussions}

For the model of the electromagnetic field coupled to the Weyl tensor of gravitational fields, the photons at low frequencies no longer propagate along the light cone in the original spacetime. That is to say, the  interactions between the electromagnetic field and the spacetime curvature lead to a dependence of the photon velocity on the motion and polarization directions, so that in some cases photons travel at speeds greater or less than unity. The  photons which travel at speeds greater or less than unity arise some difficulties for  the physics of black hole because we can not get the position of the event horizon by means of the null surface, and we can not calculate quantum entropy of the black hole by the statistical mechanics method because we do not have any knowledge of the statistical mechanics. It is interesting to note that all these difficulties can be overcome by introducing an effective spacetime in which photons propagate along the light cone.

Using this effective descriptions, we found, comparing with the results of original spacetime,  that the event horizon and Hawking temperature do not change, but the area of the event horizon and Bekenstein-Hawking entropy become different. We noted that the total entropy for this system, which is still the same as that of original spacetime, consists of two parts: one is  the Bekenstein-Hawking entropy $ \tilde{S}_H$, the other is the entropy $ \tilde{S}_C$ arising from the coupling of electromagnetic field and the Weyl tensor.  Fortunately, we also obtained effective descriptions for the Smarr
relation and first law of thermodynamics.

\section*{acknowledgments}

This work is supported by the  National Natural Science Foundation of China under Grant Nos. 11475061;  S. Chen's work was
partially supported by the National Natural Science Foundation of China under Grant No. 11275065;  and Q. Pan's work was partially
supported by the National Natural Science Foundation of China under Grant No. 11275066.


\end{document}